\documentclass[11pt]{article}
\usepackage{style}

\newcommand{\inst}[1]{\vspace{1pt} \unskip\(^{#1}\)}

\begin{document}

\title{Adding smoothing splines to the SAM model improves stock assessment}

\date{}

\author{%
  Silius M. Vandeskog\inst{1} \and
  Magne Aldrin\inst{1} \and
  Daniel Howell\inst{2} \and
  Edvin Fuglebakk\inst{2}
}

\maketitle

\noindent \inst{1} Norwegian Computing Center, P.O.Box 114 Blindern, Oslo, NO-0314, Norway \\
\inst{2} Institute of Marine Research, P.O.Box 1870 Nordnes, Bergen, NO-5817, Norway \\

\begin{abstract}
  The stock assessment model SAM contains a large number of age-dependent parameters that must be
  manually grouped together to obtain robust inference.  This can make the model selection process
  slow, non-extensive and highly subjective, while producing unrealistic looking parameter estimates
  with discrete jumps. We propose to model age-dependent SAM parameters using smoothing spline
  functions. This can lead to more smooth parameter estimates, while speeding up and making the
  model selection process more automatic and less subjective. We develop different spline models
  and compare them with already existing SAM models for a selection of 17 different fish stocks,
  using cross- and forward-validation methods. The results show that our automated spline models
  overall outcompete the officially developed SAM models. We also demonstrate how the developed
  spline models can be employed as a diagnostics tool for improving and better understanding
  properties of the officially developed SAM models.
\end{abstract}

\noindent%
\vfill

\section{Introduction}%
\label{sec:introduction}%

In recent decades, global fisheries have contributed around 90 million tonnes of fish annually into
the human food chain~\citep{FAO2024StateWorldFisheries}. The FAO estimates that the overall catch of
marine fish has fluctuated around this level for 25 years and that \(88\%\) of fish stocks are
currently fully or overexploited, reflecting the high level of exploitation involved in modern
fisheries. However, the fraction of stocks fished unsustainably is estimated to be \(37.7\%\) and
increasing. Good management, underpinned by robust stock assessment, is associated with better stock
health and productivity with fish stocks less likely to be at low stock
size~\citep{HilbornEtAl2020Effectivefisheriesmanagement}.

Where sufficient data exists, modern stock assessment are based on a paradigm of integrated
analysis: models which produce a simulated fish population which is then statistically tuned to
match data from a variety of different sources, accounting for the uncertainty in that
data~\citep{FournierEtAl1982GeneralTheoryAnalyzing}. A second paradigm is that of state-space
models, which model the state of the fish population using one or more first order difference
equations. Such models can explicitly model both the trend of the fish stock and the uncertainty
around that estimate. Assessment models which combine these approaches are now commonly used for
management of large data-rich fish stocks~\citep{Punt2023Thosewhofail}. For European stock
assessments coordinated by ICES, the dominant stock assessment model is
SAM~\citep{NielsenEtAl2014Estimationtimevarying, BergEtAl2016Accountingcorrelatedobservations} which
accounted for 46 out of the 95 stocks with data rich stock assessment models~\citep{ices2025}.

The SAM model contains a large collection of age-dependent parameters that must be estimated for
each age group of interest for any given fish stock.  The number of parameters that must be
estimated can therefore grow large for a complex fish stock with a large number of age groups and
many different age-dependent parameters to be estimated.  These parameters govern the structure of
the processes modelled (e.g.\ the fishing pressure at age) and the degree of noise in a given data
set (e.g.\ variance at age in a survey). Having too many parameters can easily lead to poor model
stability and potential difficulties during model estimation. The common approach to reduce the
number of parameters is to partition the set of all age-dependent parameters into different subsets,
and then to make all parameters inside each subset identical. A typical property of fisheries data
is that ages which are poorly sampled have higher variance (i.e.\ more noisy data) than ages which
are better sampled. This poor sampling could e.g.\ arise from small fish which are only partly
within the area of sampling or of a size where they are often able to escape through holes in the
fishing net. Conversely, older fish might be poorly sampled due to there being few individuals of
that age surviving in the population to catch. Finally, for the very oldest fish, the sample size is
likely too small for the model to be able to estimate variance at age, and combining ages to produce
a pooled variance will be necessary for model fitting. Thus, an age-dependent variance parameter
might, e.g., have unique values for the two smallest age groups, identical values for the four
oldest age groups, and identical values for all remaining age groups. However, the ages at which
these changes occur will vary according to fish life history and the behaviour of the fleet or
scientific survey. Therefore, it is often neither obvious nor intuitive exactly how the different
age parameters (and especially the variance parameters) relate to each other. The partitioning
procedure therefore reduces the number of parameters, but it can also be rather ad hoc and the
resulting model may be significantly mis-specified.

To the best of our knowledge, no automated procedures exist for selecting a suitable partition of
the age-dependent parameters of the SAM model. This model selection process must therefore be
performed manually. Simple combinatorics show that the number of possible partitions explodes as the
number of age groups and age-dependent parameters grow. Manual model selection procedures will
therefore tend to be non-extensive, by only comparing a small fraction of all possible
partitions. This can in turn make the model selection process slow and highly subjective. Manual
partitioning of age-dependent parameters can also provide somewhat unrealistic parameter estimates,
as parameters from neighbouring age groups either must be set equal to each other or estimated
separately. As aging is a continuous process, it might seem more realistic to assume that the
age-dependent parameters should evolve smoothly with age, instead of evolving as a piecewise
constant function with possibly large jumps between each partition.

In this paper, we propose to model age-dependent parameters of the SAM model using smoothing splines
instead of the manual partitioning method. A similar approached is proposed by
\citet{AldrinEtAl2020specificationdatamodel} and \citet{AldrinEtAl2021Caveatsestimatingnatural}, who
model age-dependent parameters in simplified SAM models using second-degree polynomials. We take
this one step further, by using more generalised smooth functions, and using the full version of the
official SAM model. We also develop code that makes it easy for others to adopt or extend our
methods for improved stock assessment with SAM. Our spline models guarantee that all age-dependent
parameters can obtain different values, which can lead to more biologically plausible parameter
estimates. They can also reduce the effective number of model parameters, by constraining the
parameters to change smoothly as a function of age. Modelling the age-dependent parameters with
smoothing splines also allows for a faster and more automated model selection procedure, that
required less manual fiddling with parameter configurations. This can in turn lead to a less
subjective model selection process. The developed smoothing spline model may also be used as a
diagnostics tool for improving the standard model selection procedure, for those who still want to
rely on the manual partitioning method. Initial parameter estimates, for learning which age groups
that should be given equal or different parameters, can be produced in a matter of seconds by
fitting splines to all age-dependent parameters. The model can also be used as a diagnostic for
examining the robustness of a ``standard'' SAM model with manually selected parameter partitions. If
the spline model agrees with the overall parameter estimates of the standard model, this may hint at
a robust model fit. On the other hand, if the spline model produces considerably different estimates
for certain age-dependent parameters, this may hint at a lack of model robustness, and provide
relevant information to a modeller about which parameters to focus on during the model selection
process.

The remainder of the manuscript is organised as follows: Section~\ref{sec:model} describes the main
parts of the SAM model that are relevant for our work. Section~\ref{sec:method} describes our
proposed method for estimating age-dependent model parameters with smoothing splines. In
Section~\ref{sec:case-study}, we develop new models based on the spline methodology, and we compare
these with the official SAM models for 17 different fish stocks, using both cross-validation and
forward-validation methods. Finally, in Section~\ref{sec:discussion}, we conclude the manuscript
with a short discussion.

\section{Model}%
\label{sec:model}

For a specific fish stock, we denote the number of fish in age group \(a\) at the start of year
\(y\) as \(N_{a, y}\). We want to estimate \(N_{a, y}\) for the \(A\) age groups
\(a = 1, 2, \ldots, A\). The final age group, \(A\), is known as the plus group. It contains all age
groups that are larger than \(A-1\). We use the SAM model, implemented in the \texttt{R} package
\texttt{stockassessment}~\citep{NielsenEtAl2014Estimationtimevarying,
  BergEtAl2016Accountingcorrelatedobservations}, to estimate \(N_{a, y}\). The SAM model describes
the population dynamics of a fish stock as
\begin{equation}
  \label{eq:population-model}
  \begin{aligned}
    N_{a + 1, y + 1} &= N_{a, y} \exp\left(- \{F_{a, y} + M_{a, y}\}\right) \text{ for } 1 < a < A, \\
    N_{A, y + 1} &= N_{A - 1, y} \exp\left(- \{F_{A - 1, y} + M_{A - 1, y}\}\right) +
    N_{A, y} \exp\left(- \{F_{A, y} + M_{A, y}\}\right), \\
    N_{1, y + 1} &= R_{y + 1}, \\
  \end{aligned}
\end{equation}
where \(F_{a, y}\) is the fishing mortality rate and \(M_{a, y}\) is the natural mortality rate, the
latter commonly assumed to be a known value. Recruitment is controlled by the random variable
\(R_{y + 1}\), which can take many different forms. Two popular alternatives consist of setting
\(R_{y + 1}\) equal to a random function of \(R_y\) or a random function of last years stock
spawning biomass (SSB), respectively~\citep[e.g.][]{NielsenEtAl2014Estimationtimevarying,
  BergEtAl2016Accountingcorrelatedobservations}. Multiple different recruitment options are
available in the SAM model. In practice, fish stock data sets often start at age \(0\) or at some
other age larger than \(1\). However, for notational convenience, we always assume that the smallest
age in a fish stock data set is \(1\) in this Section.

To estimate the unknowns in the population equations~\eqref{eq:population-model}, one must link the
population equations to some kind of observed data. The SAM model contains observation equations for
many different types of stock assessment data, but we focus on the two main data sources used for
stock assessment modelling, namely catch data and survey index data. The exact catch-at-age,
\(C_{a, y}\), in year \(y\) is clearly unknown. However, an estimate for \(C_{a, y}\) can be
produced, e.g.\ with the Estimating Catch-at-Age (ECA)
model~\citep{HirstEtAl2012Bayesianmodellingframework}. We denote this estimate as \(\hat C_{a,
  y}\). In the SAM model, the logarithm of the catch-at-age estimate is modelled as
\begin{equation}
  \log \hat C_{a, y} = \log \left(\frac{F_{a, y}}{F_{a, y} + M_{a, y}} (1 - e^{-(F_{a, y} + M_{a,
        y})}) N_{a, y} \right) + \gamma_{a, y},
\end{equation}
where \(\gamma_{a, y}\) is a Gaussian random variable with zero mean and age-specific variance
\(\sigma^2_{a}\).

Multiple different types of surveys may have been conducted for a single fish stock.  We sort the
different survey types based on the year they first started, and denote the survey index from the
\(j\)th survey type as \(\hat I^{(j)}_{a, y}(d)\), where \(d\) is the number of days into the year where
the survey is conducted. In the SAM model, the logarithm of the survey index is assumed to be
Gaussian distributed as
\begin{equation}
  \log \hat I^{(j)}_{a, y}(d) = \log \left(Q_{a, j} N_{a, y}(d)\right) + \varepsilon^{(j)}_{a, y},
\end{equation}
where the age- and survey-specific catchability parameter \(Q_{a, j}\) provides a link between the
survey index and the true fish population at day \(d\) of year \(y\). Additionally,
\(\varepsilon^{(j)}_{a, y}\) is Gaussian distributed with zero mean and variance
\(\omega^2_{a, j}\).

\section{Method}%
\label{sec:method}

\begin{figure*}
  \centering
  \includegraphics[width=.99\linewidth]{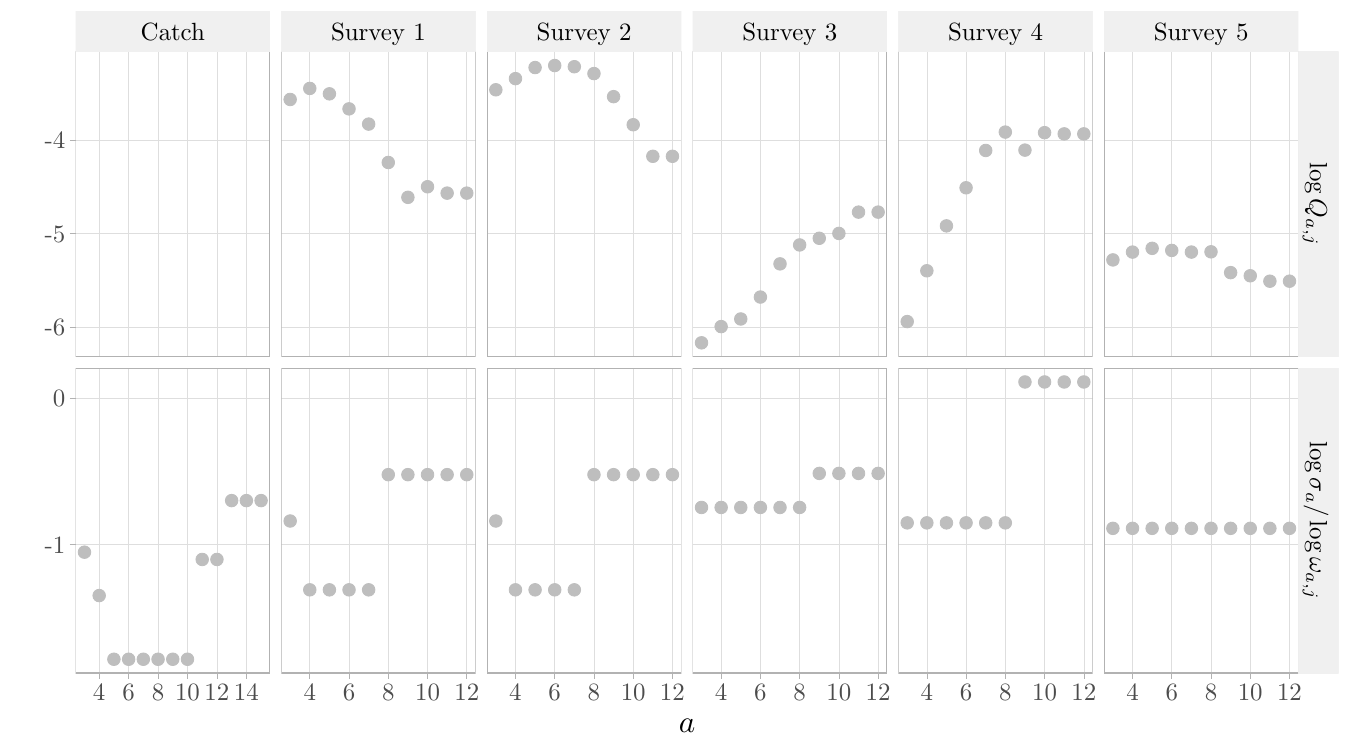}
  \caption{Age-dependent parameter estimates from the official SAM model for cod in the North-East
    Arctic.  The upper rows for each fish stock contains estimates of \(\log Q_{a, j}\) for all the
    different surveys.  The lower rows contains estimates of \(\log \sigma_a\) in the first column,
    and \(\log \omega_{a, j}\) in the remaining columns.}%
  \label{fig:cod_params}
\end{figure*}

The unknowns \(Q_{a, j}\), \(\omega_{a, j}^2\), \(\sigma^2_a\), \(F_{a, y}\) and \(N_{a, y}\) need
to be estimated for all age groups. Some of the parameters must also be estimated for different
years or different survey types. The SAM model also includes other optional model components with
even more age-dependent parameters, such as e.g.\ the density-dependent catchability power
parameters known as \texttt{Qpow} in the SAM model. As described in Section~\ref{sec:introduction},
the number of parameters therefore grows quickly with the number of age groups, and it is common to
partition the set of all age-dependent parameters into different subsets, in which all parameters
are assumed to be equal. Figure~\ref{fig:cod_params} displays an example of such a partition, for
the parameters \(Q_{a, j}\), \(\omega_{a, j}^2\) and \(\sigma^2_a\), used for modelling cod in the
North-East Arctic. For the catch variances, the number of unique parameters is reduced from \(13\)
to \(5\), while for survey 5, the number of variance parameters is reduced all the way from \(10\)
to \(1\). For a fish stock with \(A\) different age groups, there are \(2^{A-1}\) different
partitions of the age-dependent parameters. To the best of our knowledge, no automated or
standardised procedures have been developed for creating optimal parameter partitions. The
partitioned parameters tend to look somewhat too discrete, with neighbouring parameters that
either are identical or far away from each other.  This can be seen in the variance parameters in
Figure~\ref{fig:cod_params}. The estimates in the three leftmost variance subplots imply that the
observation variances should behave somewhat like skewed parabolas, which start by smoothly
decreasing until some minimum, and then starts increasing after that. However, the partitioned
parameters instead tend to produce a piecewise constant functions with large jumps between each
subset. This behaviour is highly unrealistic. Ageing is a continuous process, and most properties
related to ageing should therefore be expected to change smoothly as a function of age.

We propose to improve the model-selection process by imposing a smooth structure on the
age-dependent parameters. Thus, instead of estimating the parameters
\((\theta_{1}, \theta_{2}, \ldots, \theta_{A})\), we estimate the parameters of a smooth parametric
function \(f_\theta(a)\) such that \(\theta_a = f_\theta(a)\) for all \(a = 1, 2, \ldots, A\). This
can speed up the model selection process, since a modeller only needs to consider a few different
smooth functions, instead of all \(2^{A - 1}\) possible partitions for each age-dependent
parameter. It can also make the model selection process less subjective, while making the
subjectivity of the modeller less tangible. When a modeller has chosen a partition of \(\theta_a\),
it might be difficult to understand why that specific partition was chosen. However, when a modeller
chooses a smooth function \(f_\theta\), all their assumptions are incorporated into that function.
A model critic can then look at the function \(f_\theta\), find that it e.g.\ allows for too abrupt
changes, and modify it so it better suits their own assumptions.

We model the parameters \(Q_{a, j}\), \(\omega^2_{a, j}\) and \(\sigma^2_a\) using smoothing
splines. Smoothing of \(F_{a, y}\) is outside the scope of this paper. This is because the fishing
mortality rates are assumed to be correlated across years, which complicates any smoothing
procedures. A spline function, \(f(a)\), is a linear combination of polynomials,
\begin{equation}
  f(a) = \sum_{i = 1}^n \beta_i \phi_i(a),
\end{equation}
where the \(n\) parameters \(\bm \beta = (\beta_1, \ldots, \beta_n)^\top\) describe the weights of
each of the \(n\) polynomials \(\phi_1, \ldots, \phi_n\), also known as the basis functions.  Using
matrix notation, the spline function can be written as \(f(a) = \bm x^\top(a) \bm \beta\), where the
vector \(\bm x(a)\) contains all \(n\) basis splines, evaluated at \(a\). Furthermore, the vector
\(\bm f = (f(1), f(2), \ldots, f(A))^\top\) can be written as \(\bm f = \bm X \bm \beta\), where the
matrix \(\bm X = (\bm x(1), \bm x(2), \ldots, \bm x(A))^\top\) contains all basis splines evaluated
at all age groups.

Splines containing a large number of basis functions tend to be highly flexible and can easily
overfit to the training data. It is therefore common to regularise the splines in some way by adding
penalty terms to the likelihood function~\citep[e.g.][]{Wood2004StableEfficientMultiple}. These
penalty terms can then be designed to penalise abrupt changes, too much wiggliness, too large
absolute values, and many other properties of interest. We assume that the differences between age
group 1 and 2 should be more considerable than the differences between e.g.\ age group 9 and
10. Therefore, we implement the smoothing splines as functions of \(\log(a + 1)\) instead of
functions of \(a\). In this log-space, the distance between age group 1 and 2 is as large as the
distance between age group 3 and 5 and between age group 5 and 8. To constrain the spline parameters
\(\bm \beta\), our penalised log-likelihood takes the relatively common form
\begin{equation}
  \label{eq:penalised_likelihood}
  \mathcal L_{\bm \lambda}(\bm \theta, \bm \beta) = \ell(\bm \theta, \bm \beta) - \frac12 \sum_{i =
    1}^{M} \lambda_i \bm \beta^\top \bm S_i \bm \beta,
\end{equation}
where \(\ell\) is the original log-likelihood without any penalty terms, \(\bm \theta\) is a vector
of all remaining model parameters and \(\bm S_1, \ldots, \bm S_M\) is a set of \(M\) penalty
matrices, describing which properties of \(\bm \beta\) we wish to penalise. Finally,
\(\bm \lambda = (\lambda_1, \ldots \lambda_M)\) is a vector of \(M\) penalty parameters, describing
how much weight that should be given to each of the penalty matrices. \(M\) is typically equal to
either \(1\) or \(2\), as a large number of penalty parameters leads to overly complicated models.

We use the \texttt{mgcv} package~\citep{Wood2017GeneralizedAdditiveModels} in \texttt{R} for
designing and computing the penalty matrices and the values of the basis functions at all age groups
of interest. \texttt{mgcv} makes it easy to design a large variety of different spline functions
with different types of penalties. Note, that our developed SAM model is not constrained to using
\texttt{mgcv}. It is method agnostic, in the sense that it only requires the matrices \(\bm X\) and
\(\bm S_1, \ldots, \bm S_M\) as input. To avoid subjective and time-demanding decisions about the
number of basis functions during the model selection process, we set the number of basis functions
equal to the number of age groups. Without any penalisation this would heavily increase the number
of model parameters, and likely lead to considerable overfitting. However, given reasonable spline
penalties, the parameters will be constrained in a way that reduces the wiggliness, and thus the
effective number of parameters, to a suitable level for any data set in question. In
Section~\ref{sec:case-study}, we choose two specific spline functions from \texttt{mgcv} for
comparing our model framework against a selection of official SAM models. More details about these
two spline functions are given there.

The smoothing penalty parameters \(\bm \lambda\) should be large enough to reduce overfitting, but
not so large that the splines loose all their flexibility. Optimal values for \(\bm \lambda\) cannot
be estimated by maximising the penalised log-likelihood~\eqref{eq:penalised_likelihood}, as this
would result in a penalty of zero. From a Bayesian perspective, the penalised likelihood has the
same mode and overall shape as a posterior distribution for \(\bm \beta\) where the (improper) prior
distribution of \(\bm \beta\) is Gaussian with zero mean and ``inverse covariance matrix''
\(\sum_{i = 1}^M \lambda_i \bm S_i\)
\citep[e.g.][]{KimeldorfEtAl1970CorrespondenceBayesianEstimation, Wahba1978ImproperPriorsSpline,
  WoodEtAl2016SmoothingParameterModel}. Inspired by e.g.\ \citet{Wood2010FastStableRestricted}
and \citet{WoodEtAl2016SmoothingParameterModel}, we use this Bayesian perspective to estimate the
model parameters and smoothing penalty parameters by maximising the logarithm of the ``posterior''
probability density function,
\begin{equation}
  \label{eq:posterior}
  \ell(\bm \theta, \bm \beta) + \log \pi_{\bm \lambda}(\bm \beta),
\end{equation}
where \(\pi_{\bm \lambda}(\bm \beta)\) is the probability density function of the Gaussian prior for
\(\bm \beta\). More details about the parameter estimation is provided in Appendix~\ref{app:penalty}.

\section{Case study}%
\label{sec:case-study}

\subsection{Method}

We compare the developed smoothing spline SAM with official SAMs for a large collection of stock
assessment data sets. The webpage \url{stockassessment.org} contains stock assessment data for a
large variety of fish stocks, together with official SAMs, where partitions for the age-dependent
parameters already have been chosen. We download stock data together with their corresponding
official models for 17 different stocks. These are listed in Table~\ref{tab:stocks}. The 17 stock
data sets were chosen to represent a wide spread of different areas and fish stocks, given the
constraint that we only want to model stock data sets where the corresponding SAM model is described
as the final version on \url{stockassessment.org}. For each stock data set, we compare the
official SAMs against competing smoothing spline SAMs, by examining parameter estimates, population
estimates and log-likelihoods. Model comparison is also performed using cross-validation and
forward-validation studies.

\begin{table*}
  \centering
  \caption{All fish stock data used for the model evaluation. The columns \textit{Years catch/survey}
    contain all years with available catch/survey data, \textit{Ages} contains the minimum and maximum
    ages in the data set and \textit{Data source} contains the names used to identify the data sets at
    \url{stockassessment.org}.
  }%
  \label{tab:stocks}
  \resizebox{\textwidth}{!}{%
  \begin{tabular}{llllll}
    \toprule
    Fish stock & Area & Years catch & Years survey & Ages & Data source \\
    \midrule
    Cod & Baltic Sea & 1985 - 2021 & 1996 - 2022 & 0 - 7 & WBcod22Fsq \\
    Herring & Baltic Sea & 1977 - 2021 & 1999 - 2021 & 0 - 8 & GoR\_BP\_v2.2.3qF\_s \\
    Plaice & Baltic Sea & 1999 - 2022 & 1999 - 2022 & 1 - 7 & ple.27.21-23\_WGBFAS\_2023\_ALT\_v1 \\
    Cod & Celtic Sea & 1980 - 2022 & 2002 - 2022 & 0 - 7 & Cod\_7ek\_2023 \\
    Haddock & Celtic Sea & 1993 - 2022 & 2003 - 2022 & 0 - 8 & HAD7bk\_2023\_final \\
    Plaice & Celtic Sea & 1989 - 2021 & 1989 - 2021 & 1 - 10 & Ple.7fg.2022.main \\
    Whiting & Celtic Sea & 1999 - 2022 & 2000 - 2022 & 0 - 7 & whg.7b-ce-k\_WGCSE22\_RevRec\_2023 \\
    Haddock & Faroe Plateau & 1957 - 2023 & 1994 - 2023 & 1 - 10 & NWWG2023\_faroehaddock \\
    Ling & Faroe Plateau & 1996 - 2022 & 1996 - 2022 & 3 - 12 & lin.27.5b\_wgdeep2023\_final \\
    Saithe & Faroe Plateau & 1961 - 2023 & 1994 - 2023 & 3 - 15 & fsaithe-NWWG-2023 \\
    Cod & North-East Arctic & 1946 - 2022 & 1981 - 2023 & 3 - 15 & NEA\_cod\_2023\_final\_run \\
    Saithe & North-East Arctic & 1960 - 2021 & 1994 - 2021 & 3 - 12 & NEAsaithe\_2022\_v3 \\
    Haddock & North Sea & 1972 - 2022 & 1983 - 2023 & 0 - 8 & NShaddock\_WGNSSK2023\_Run1 \\
    Plaice & North Sea & 1957 - 2020 & 1970 - 2020 & 1 - 10 & plaice\_final\_10fix \\
    Sole & North Sea & 1984 - 2021 & 1987 - 2021 & 1 - 9 & Sole20\_24\_2022vs21 \\
    Whiting & North Sea & 1978 - 2022 & 1983 - 2023 & 0 - 6 & NSwhiting\_2023 \\
    Blue whiting & Widely distributed & 1981 - 2023 & 2004 - 2023 & 1 - 10 & BW-2023 \\
    \bottomrule
  \end{tabular}
  }
\end{table*}

We develop two competing smoothing spline models, using two different spline functions from
\texttt{mgcv}. The first model, denoted \textit{Spline1}, uses a shrinkage version of a cubic
regression spline (denoted \texttt{cs} in \texttt{mgcv}). The second model, denoted
\textit{Spline2}, uses a B-spline basis (denoted \texttt{bs} in \texttt{mgcv}) of order 3. Both
models penalise the second derivative of the spline functions. For an improved comparison basis, we
also evaluate a fourth model, denoted the \textit{Maximal} model. In this model, the
age-dependent parameters are not constrained in any way. This results in a highly flexible, but less
robust model, with a large amount of parameters. The four competing models are listed and described
in Table~\ref{tab:models}. Note that, while all four models use different configurations for the
parameters \(Q_{a, j}\), \(\omega_{a, j}^2\) and \(\sigma_a^2\), all other parameter configurations
are kept unchanged from that of the official model. We use the default initial values in SAM for all
four models. These are \(\log Q_{a, j} = -5\) and \(\log \omega_{a, j} = \log \sigma_a = -0.35\). We
use an initial value of \(0\) for the logarithms of the smoothing penalty parameters.

\begin{table*}
  \centering
  \caption{Names and descriptions of the four competing models.}%
  \label{tab:models}
  \begin{tabular}{p{.1\textwidth}p{.85\textwidth}}
    \toprule
    Model & Description \\
    \midrule
    Official & The official model, with manually selected partitions of the age-dependent parameters, retrieved from
               \url{stockassessment.org}. \\
    Spline1 & \(Q_{a, j}\), \(\omega_{a, j}^2\) and \(\sigma_a^2\) are modelled with shrinkage versions of a cubic
              regression spline (\texttt{cs}) that penalises large absolute values of the second
              derivatives. \(\omega_{a, j}^2\) and \(\sigma_a^2\) are given one common penalty
              parameter, that is the same for all values of \(a\) and \(j\).
              \(Q_{a, j}\) is given its own penalty parameter, that is the same for all values of
              \(a\) and \(j\). This results in a total
              of \(2\) penalty parameters for the entire model. \\
    Spline2 & \(Q_{a, j}\), \(\omega_{a, j}^2\) and \(\sigma_a^2\) are modelled with splines based
              on B-spline bases of order 3 (\texttt{bs}) that penalises large absolute values of the second
              derivatives. \(\omega_{a, j}^2\) and \(\sigma_a^2\) are given  one common penalty
              parameter, that is the same for all values of \(a\) and \(j\).
              \(Q_{a, j}\) is given its own penalty parameter, that is the same for all values of
              \(a\) and \(j\). This results in a total
              of \(2\) penalty parameters for the entire model. \\
    Maximal & \(Q_{a, j}\), \(\omega_{a, j}^2\) and \(\sigma_a^2\) are given independent
                 parameters for all age groups of interest. \\
    \bottomrule
  \end{tabular}
\end{table*}

We evaluate the models' performances using cross-validation and forward-validation. This is achieved
by removing one or more years of data, fitting the competing models to the remaining data, and then
predicting properties of the removed data. The best way of evaluating model performance would be to
compare predicted populations with the true populations. This is clearly not possible, since the
true populations are unknown. Instead, we evaluate the models by predicting catch-at-age data and
survey indices, and comparing these to the true values. Our hope is that the model that is best at
predicting catch-at-age data and survey indices also is the model that is best at estimating the
true populations.  Within SAM, one can easily estimate unknown catch-at-age data and survey indices
for some year \(y\), by setting all the observations from that year equal to \texttt{NA}.

Forward-validation is performed by fitting a model to all data from all years less than some year
\(y\) for a given fish stock data set.  Then, we predict catch-at-age and survey indices for year
\(y\) using the fitted model. This can be unstable if one sets \(y\) low enough that only a few
years of data are available. We therefore restrict the forward-validation so that \(y\) must be in
the last third of the available years for each fish stock data set. Several data sets contain survey
indices that have only been reported for a few years. Different values of \(y\) might therefore lead
to data sets containing only one or two years of data from a specific survey type. This can make it
hard to reliably estimate the survey parameters for that survey type. We therefore also filter out
data from any type of survey with less than five years of data before \(y\). Cross-validation is
achieved by only removing data from a single year, and predicting catches and survey indices for
that year. This is more stable, since less data is removed. We therefore perform cross-validation
for all available years except the first, and we never remove all the data from an entire type of
survey.

The cross/forward-validation predictions are evaluated by computing the root mean squared error
(RMSE) between predicted and observed catch-at-age data and survey indices. Catch-at-age RMSE values
are computed by summing over the squares of the differences between \(\tilde C_{a, y}\) and
\(\hat C_{a, y}\) for all age groups \(a\) and all leave-out years \(y\), where \(\tilde C_{a, y}\)
is predicted catch-at-age data obtained using the described cross-validation or forward-validation
procedures. Survey index RMSE is computed similarly, where we sum over the squares of the
differences between \(\tilde I^{(j)}_{a, y}\) and \(\hat I^{(j)}_{a, y}\) for all age groups \(a\),
all years of interest \(y\) and all available survey types, where \(\tilde I^{(j)}_{a, y}\) is a
predicted survey index obtained using the described cross-validation or forward-validation
procedures. During the forward-validation, we also perform a conditional catch-at-age prediction,
where we predict catch-at-age values for year \(y + 1\) given that we know the total biomass of all
catches for that year. One can think of this as predicting next years catch-at-age values given that
the fishing quota for next year has been set, which is a highly realistic scenario. Conditional
prediction is performed using the \texttt{forecast()} function in SAM. Having predicted conditional
catch-at-age values, we then compute RMSE values just as we did for the unconditional catch-at-age
values.

\subsection{Results}

\begin{table}
  \centering
  \caption{Convergence statistics for the cross/forward-validation.}%
  \label{tab:convergence}
  \begin{tabular}{lrll}
    \toprule
    Model & \multicolumn{3}{c}{Convergence} \\
    \midrule
    Official & \(1009 / 1015\) & \(\approx\) & \(99.4\%\) \\
    Spline1 & \(970 / 1015\) & \(\approx\)& \(95.6\%\) \\
    Spline2 & \(996 / 1015\) & \(\approx\) & \(98.1\%\) \\
    Maximal & \(855 / 1015\) & \(\approx\) & \(84.2\%\) \\
    \midrule
    All & \(810 / 1015\) & \(\approx\) & \(79.8\%\) \\
    \bottomrule
  \end{tabular}
\end{table}

In total, we fit each of the four SAM models to \(1015\) different data sets during the
cross/forward-validation. As some of the models are less robust than the others, and as no tweaking
of initial values is performed, it is expected that not all four models will be able to converge
successfully at first try for all the \(1015\) different data sets. Table~\ref{tab:convergence}
displays the total number of successful convergences for each of the four models. As expected, the
maximal model is the least robust, as it contains the largest number of parameters. Furthermore, the
official model is the most robust, as it has been designed specifically for each fish stock data set
by a panel of experts. Interestingly, the convergence rate of the spline2 model is only slightly
lower than that of the official model. This is a promising result, as the spline models do not
contain any prior knowledge about the shapes of \(Q_{a, j}\), \(\omega_{a, j}^2\) and
\(\sigma_a^2\), except the assumptions that these should behave somewhat smoothly as a function of
\(a\). No tweaking of initial values has been performed either. However, note that the remaining
parameter configurations of the spline models are the same as in the official model. The bottom row
shows that all four models successfully converged for \(810\) of the \(1015\) data sets. We only
compute and compare RMSE values for these \(810\) data sets, to make the model comparison as fair as
possible.

\begin{figure*}
  \centering
  \includegraphics[width=.99\linewidth]{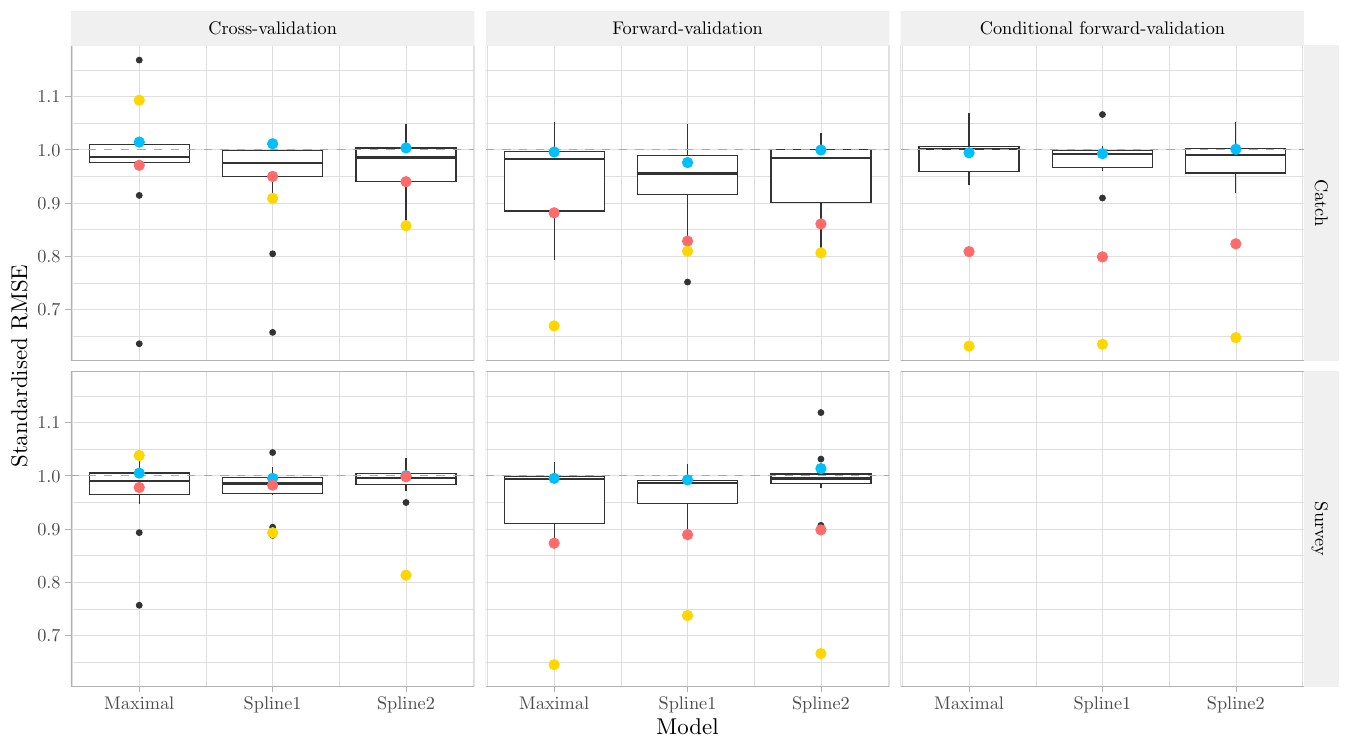}
  \caption{Box-plots displaying standardised RMSE values for the 17 different fish stock data sets
    in Table~\ref{tab:stocks}. The boxes span from the \(25\) percentile to the \(75\) percentile of
    the 17 standardised RMSE values, and the dark horisontal line within each box displays the
    standardised RMSE median. The leftmost column displays standardised RMSE values for catch-at-age
    data and survey indices from the cross-validation. The middle column displays the same for the
    forward-validation. The rightmost column displays standardised RMSE values for the conditional
    catch-at-age data from the forward-validation procedure. The blue, red and yellow dots
    correspond to the fish stocks in subplot A, B and C of Figure~\ref{fig:params},
    respectively. The blue dot in the bottom left subplot is hidden behind the red dot.}%
  \label{fig:evaluation}
\end{figure*}

For easier comparisons between the official model and its three competing models, we standardise the
RMSE values by dividing all RMSE values from the competing models on their corresponding RMSE values
from the official model. A standardised RMSE greater than \(1\) means that the official model
outperformed the competing model, while a standardised RMSE less than \(1\) means the
opposite. Figure~\ref{fig:evaluation} displays box-plots for the standardised RMSE values between
the official model and the other three competing models, for all of the 17 fish stock data
sets. Conditional prediction was not performed for Haddock or Saithe from the Faroe Plateau, as
these stock data sets contained missing observations for at least one age group in almost every year
used for forward-validation, which makes it impossible to condition on the total biomass of last
years catches. Figure~\ref{fig:evaluation} show that all three models achieve standardised RMSE
values that are both greater than \(1\) and less than \(1\) for all the different evaluation
criteria. However, the main trend seems to be that all three of the competing models slightly
outcompete the official model for most of the fish stock data sets. The spline1 model seems to
perform the best overall, with respect to RMSE, as it achieves a standardised RMSE below 1 for the
majority of the fish stocks in each of the five sub-plots. However, the spline2 model and the
maximal model also seem to outperform the official model in most of the five subplots.

\begin{figure*}
  \centering
  \includegraphics[width=.99\linewidth]{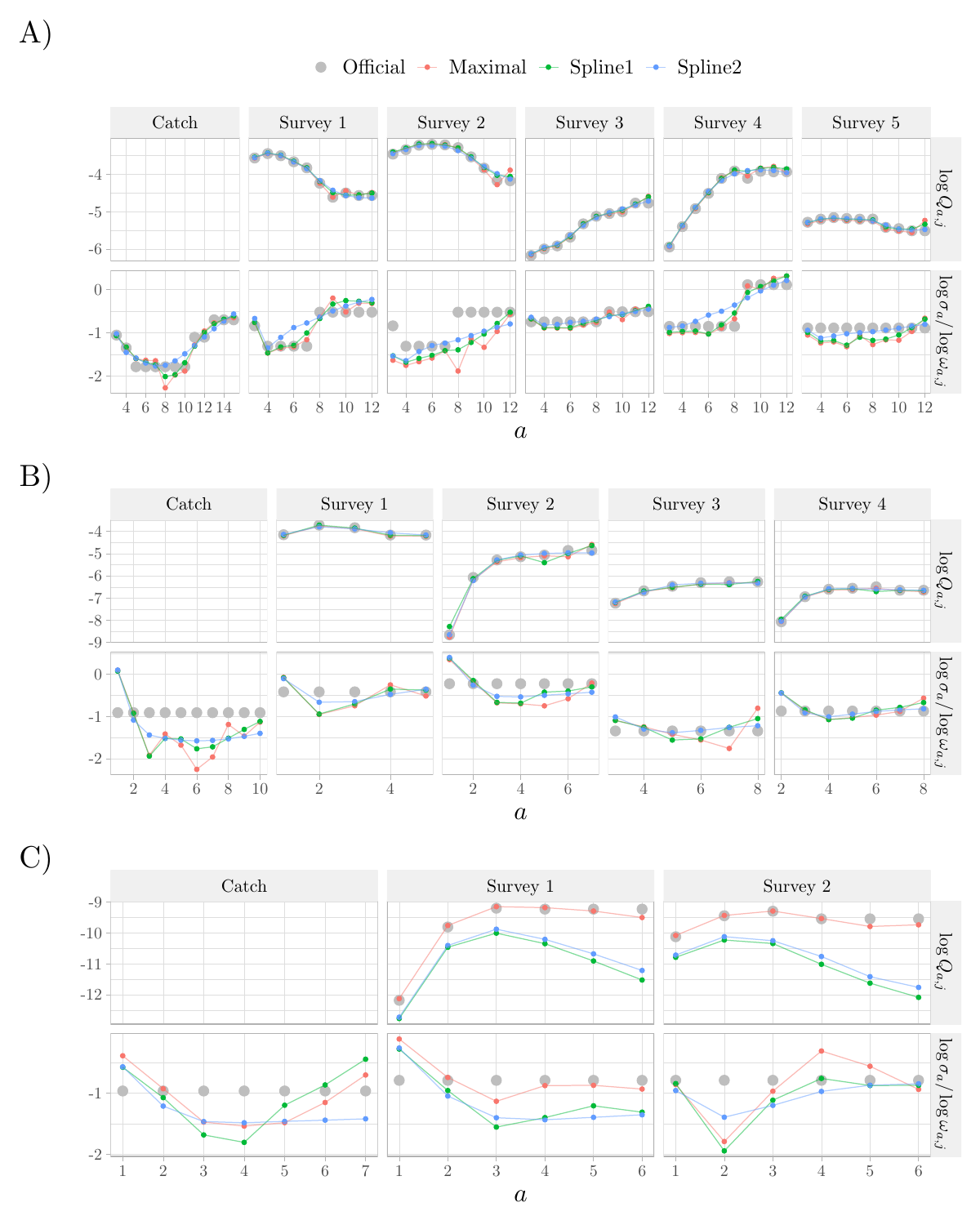}
  \caption{Age-dependent parameter estimates for three different fish stocks: A) Cod in the
    North-East Arctic, B) Plaice in the Celtic Sea and West of Scotland, C) Plaice in the Baltic
    Sea. The upper rows for each fish stock contains estimates of \(\log Q_{a, j}\) for all the
    different surveys. The lower rows contains estimates of \(\log \sigma_a\) in the first column,
    and \(\log \omega_{a, j}\) in the remaining columns.}%
  \label{fig:params}
\end{figure*}

We evaluate additional properties of the model fits by examining and comparing parameter estimates
for each of the 17 fish stocks. Estimated values of \(\log Q_{a, j}\), \(\log \omega_{a, j}\) and
\(\log \sigma_a\) for three different fish stocks are given in Figure~\ref{fig:params}. Subplot A
displays estimated parameters for cod in the North-East Arctic, in which the official model
parameters are also displayed in Figure~\ref{fig:cod_params}. All four competing models seem to
agree well about the catchability (\(Q_{a, j}\)) estimates. The models also more-or-less agree about
the variance estimates. The official model partition contains several subsets with 4-6 different
parameters, while the other three models are less constrained, and therefore tend to vary more as a
function of age. Still, the model differences are small, overall. The largest differences occur for
\(\omega_{a, 2}\), where the three competing models produce smaller estimates. We are unsure why
this happens. As shown by the blue dots in Figure~\ref{fig:evaluation}, the cross/forward-validation
also finds that all four models perform approximately equally well. However, the time and knowledge
required to develop the official model is considerably larger than what is required for the other
three models.

Subplot B of Figure~\ref{fig:params} displays estimated parameters for plaice in the Celtic Sea and
West of Scotland. All four models agree about the catchability estimates, while the differences are
larger for the variance parameters. All the official variance parameters are age-independent, while
the more flexible models claim that the variance should drop quickly for the first age groups, and
then stay constant or slowly increase for the larger age groups. This added flexibility seems to
lead to a considerable model improvement, because the official model is outperformed by the other
three models in both the cross-validation and the forward-validation, as shown by the red dots in
Figure~\ref{fig:evaluation}.

\begin{figure*}
  \centering
  \includegraphics[width=.99\linewidth]{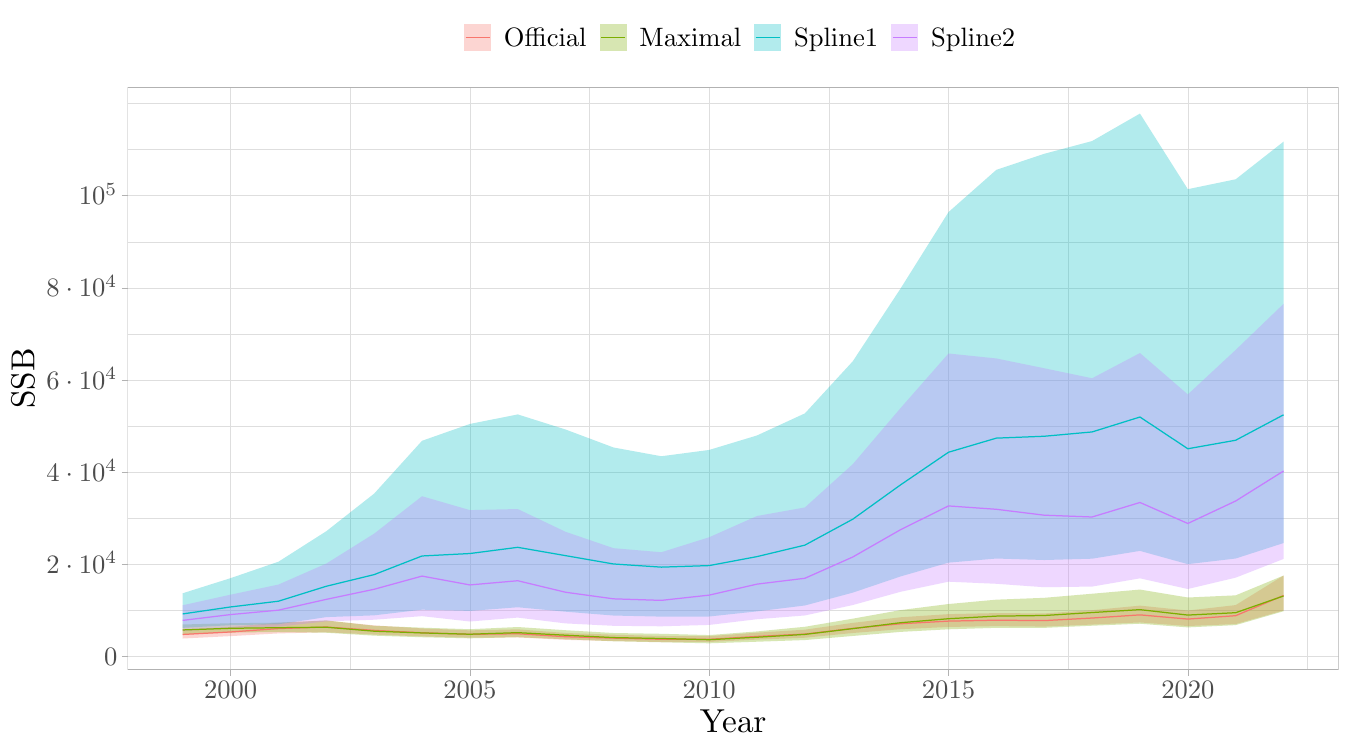}
  \caption{SSB for plaice in the Baltic Sea. The solid lines display the yearly estimated SSB
    values for each SAM model, while the shaded areas display estimated \(95\%\) confidence
    intervals for the yearly SSB values.}%
  \label{fig:plaice-ssb}
\end{figure*}

Subplot C of Figure~\ref{fig:params} displays estimated parameters for plaice in the Baltic
Sea. This is the only one of the 17 fish stocks where we find considerable differences in the
catchability estimates between different models. Since \(\log \hat I_{a, y}^{(j)}(d)\) is a random
variable with mean \(\log \left(Q_{a, j} N_{a, y}(d)\right)\), a decrease in \(Q_{a, j}\) must also
lead to an increase in \(N_{a, y}(d)\), or a large increase in \(\omega_{a, j}\). As expected,
\(N_{a, y}(d)\) is therefore estimated to be considerably larger for the spline models than for the
official and maximal models. Figure~\ref{fig:plaice-ssb} displays the estimated stock spawning
biomass (SSB) for each model. The spline models produce SSB values more than twice as large as for
the two other models. Since the true SSB is unknown, it is impossible to say which model that is
most correct. Still, it is concerning that such a small change to the parameter configuration can
lead to so considerable changes in the estimated fish populations and SSB values.  The changes in
\(Q_{a, j}\) seem to happen because the official model assumes identical catchability for the three
oldest age groups, while the spline models believe that the catchability should decrease for the
oldest age groups. If we modify the official model so the thee largest age groups can obtain
different catchabilities, we find that it changes considerably and almost perfectly agrees with
the spline2 model. It is unclear why the maximal model does not agree with the spline
models. However, further examinations show that the maximal model fit is quite unstable, and
small changes to its initial parameters can cause it to agree with the spline models.  The yellow
dots in Figure~\ref{fig:evaluation} show that the official model is considerably outperformed by the
three competing models for most of the evaluation metrics, but that the maximal model performs
worse during the cross-validation. This might be caused by the instability of the maximal model.

Figure~\ref{fig:params} demonstrates the power of the splines models and the maximal model as
diagnostic tools, not only as stand-alone SAM models. In subplot A, all models mostly agreed about
the different parameter estimates, except from the estimates for \(\omega_{a, 2}\), which were
considerably smaller for the maximal and the spline models. It is interesting to discover that
different models can obtain different values for a single variance parameter, while all other
variance parameters are equal. A modeller presented with this information might therefore spend more
time on examining the parameter \(\omega_{a, 2}\) specifically, by focusing on its potential
dependencies with other parameters in the SAM-model, and how different configurations or smoothness
assumptions for \(\omega_{a, 2}\) can affect the overall model fit and performance. Similarly, in
subplot B, the maximal model and the spline models all estimate that the variance parameters should
be large for the youngest age groups and then decrease and flatten out as the age increases. Knowing
this, a modeller might focus more on testing parameter configurations where the youngest age groups
are given different variance parameters than the oldest age groups. In subplot C, we find the same
variance parameter patterns as in subplot B. In addition, we find that small changes in parameter
configurations can yield completely different population estimates. Clearly, this should be of
interest to the modellers that created the official model. It might be that further investigations
show that the official population estimates are the best ones, but one should probably investigate
further why these large differences appear, and if that has any consequences for how the official
model should be treated and/or further developed.

\section{Discussion}%
\label{sec:discussion}

We propose modelling age-dependent parameters of the SAM model with smoothing splines as a guide to
model development in stock assessment. This can lead to more realistic looking parameter estimates,
as it reduces the discrete behaviour of the standard parameter partitioning method. Our method also
makes it possible to automate and speed up model selection, and it allows for a less subjective
model selection process. Finally, the spline method can be utilised as a diagnostics tool for
suggesting initial values and starting points for partitioning the age-dependent parameters in the
standard SAM model, and for examining the robustness of a model fit by comparing it to the fits of
slightly different models.

We develop two different spline models, and compare them with the official SAM models for 17
different fish stock data sets. Cross-validation and forward-validation studies demonstrate that our
spline models often outperform the official SAM models, while simultaneously requiring considerably
less effort to be developed. Code for the two developed spline models are freely available online at
\url{https://github.com/NorskRegnesentral/SAM-spline}, and can easily be used for modelling other
fish stocks. Since the spline1 model demonstrated the highest performance, while the spline2 model
was the most robust spline model, our recommendation for modellers interested in using these models
would be to start by fitting the spline1 model to your data, and to switch to the spline2 model if
they encounter convergence issues. However, our chosen spline functions are relatively standard in
the literature, and not specifically tailored for stock assessment modelling. It might therefore be
possible to develop more specialised spline functions that both perform better and that are more
robust than our two spline models. Therefore, the available code online has been developed to make
it easy to incorporate new types of smoothing splines into the SAM model, for modellers who are
interested in experimenting with other model designs.

We also compare the official model with a maximal model, where all age groups are given separate
parameters. The maximal model is also found to slightly outperform the official model, with regard
to out-of-sample root mean squared error (RMSE) values. However, the model is less robust than its
three competing models, in the sense that it more often fails to converge. It also has a tendency to
more often produce considerably different parameter estimates when new years of data are
added. Additionally, for 5 of the 17 fish stock data sets, the maximal model produces unrealistic
parameter estimates, where the variance parameter for one specific age group suddenly becomes
several orders of magnitudes smaller than the variance parameters of all other age groups and all
other SAM models, while the neighbouring age groups are given variance parameters that are
comparable to the other SAM models (results not shown). Still, we did not attempt to tweak the
models that failed to converge, or to restart parameter estimation using different initial
values. In a setting where multiple modellers focus on achieving convergence with the maximal model
for one specific fish stock data set, it might be that convergence can be achieved easily with just
a few minor changes to the model formulation or initial values.  However, as the four models in
Section~\ref{sec:case-study} only are compared for data subsets where all four models converged,
there are more than 200 data sets that were not used for model evaluation, where the official model
converged, but the maximal did not. Since the maximal model seems to struggle for these data sets,
it might be that any tweaking of initial values to improve the convergence rate would actually
decrease model performance relative to the official model, as the two models would then have to be
compared using data that are more unfavorable towards the maximal model.  All in all, it is clear
that the maximal model is less robust than the other three SAM models. However, given its promising
performance for several of the fish stock data sets, the maximal model might still serve as a viable
alternative for many different fish stock data sets, both as a competitive SAM model and as a
diagnostics tool.

We would highlight that the model output (and hence quota advice) is sensitive to the parameter
partitioning choices, i.e.\ which age-dependent parameters that are set equal to each other. The
mis-specification identified is not a purely theoretical issue, but can undermine the model
performance. The variance estimates are used within the model to allocate weight to different data
sources during model fitting. Mis-specification of the partitioning structure can therefore have
serious implications for the stock assessment, with the model either overly tuned to noisy data or
erroneously downweighting reliable data. Producing a realistic parameter partitioning is therefore
important to ensuring that there is a robust scientific underpinning to fisheries management.

Our developed smoothing spline extension of SAM is highly general, as it works for all types of
parametric functions that can be written as a linear combination of parameters and basis functions,
and it allows all penalisation structures that can be written as a quadratic polynomial of the
parameters of interest. Thus, one can easily use our framework to model the age-dependent parameters
using e.g.\ other types of spline models, 2.\ or 3.\ degree polynomials of age, trigonometric,
logarithmic or exponential functions of age, or a combination of these. We chose to model the
age-dependent parameters using cubic regression splines and B-splines of order 3, because these are
relatively common spline models that are easy to implement using the \texttt{mgcv} package in
\texttt{R}. However, given the considerable success of our these two relatively standard spline
models, we believe that further work on designing spline functions and penalties can lead to further
improvements to both model performance and model robustness.

A different approach for more flexible modelling of the variance parameters in SAM was proposed by
\citet{BreivikEtAl2021Predictionvariancerelationstate}. They argue that the relationships between
the variances and means of \(\hat C_{a, y}\) and \(\hat I^{(j)}_{a, y}\), which they denote the
prediction-variance link, might be too constraining in the original SAM model, and they add some
additional model parameters to make these relationships more flexible. In their examples, the
modified prediction-variance links seem to remove much of the age-signals in \(\sigma^2_a\) and
\(\omega^2_{a, j}\). This modified prediction-variance link might therefore reduce the need for
spline modelling of \(\sigma^2_a\) and \(\omega^2_{a, j}\). We have not attempted to model the fish
stocks from Section~\ref{sec:case-study} using spline models and a modified prediction-variance
link, but this might be an interesting topic for future research.

In this paper, we have only implemented and tested smoothing spline models for the parameters
\(Q_{a, j}\), \(\omega_{a, j}^2\) and \(\sigma^2_a\). However, many SAM models contain additional
age-dependent parameters, and all contain the age- and year-dependent unknowns for the fishing
pressure by age over time (\(F_{a, y}\)). These parameters have a direct impact on the model in
terms of the dynamics of the modelled stock (which fish are killed each year) and through model
tuning to data (which fraction of the caught fish are of a given age). Correct specification here is
therefore extremely important for good model performance. Further work should therefore focus on
implementing the smoothing splines for any additional parameters where this makes sense.
\(F_{a, y}\) might require some additional work, as it typically is assumed to vary over time, with
a degree of autocorrelation. Furthermore, since spline modelling of other parameters than
\(Q_{a, j}\), \(\omega_{a, j}^2\) and \(\sigma^2_a\) where outside the scope of this paper, our
spline models and maximal model relied on using parameter configurations from the official model for
all other age-dependent parameters. Thus, it is not yet entirely correct that our developed models
allow for a fully automated model selection procedure. However, the model selection procedure is
fully automated for \(Q_{a, j}\), \(\omega_{a, j}^2\) and \(\sigma^2_a\), and it has the potential
of becoming fully automated if similar spline models can be implemented for all other age-dependent
parameters.

As discussed, more work might be necessary for achieving robust and fully automated model selection
for all age-dependent parameters. However, when it comes to using the spline as a diagnostics tool
for the standard SAM model, no additional work is needed. Our method is easy to use, and readily
available for anyone who are interested in stock assessment modelling. We believe that our method
can be of great help for designing SAM models and for understanding the properties of an already
developed SAM model even better.  In particular the method outlined here can be used to validate the
existing model structure, and serve as a guideline any revisions. We would therefore recommend that
it be applied whenever a model revision is being conducted.

\section*{Code and data availability}
The code used for extending the SAM model and for creating all results and figures in this paper is
freely available online at \url{https://github.com/NorskRegnesentral/SAM-spline}.
The data used in this paper is freely available online at \url{stockassessment.org}.

\section*{Conflict of interest}

The authors declare no conflict of interest.

\section*{Funding}

This work was funded by the Institute of Marine Research, Norway, through the 338 project
``Rammeavtale for statistikk og beregningsmatematikk -- Saksnr 20/03100''.

\section*{Acknowledgments}

We are grateful to Olav Breivik for many helpful discussions.

\appendix

\setcounter{section}{0}
\renewcommand{\thesection}{\Alph{section}}
\numberwithin{equation}{section} 
\numberwithin{figure}{section} 
\numberwithin{table}{section} 

\section{Estimating the smoothing penalty parameter}%
\label{app:penalty}

\citet{Wood2010FastStableRestricted} and \citet{WoodEtAl2016SmoothingParameterModel} rely on a
Bayesian perspective, described in Section~\ref{sec:model}, for estimating model parameters and
smoothing penalty parameters by maximising the logarithm of the ``posterior'' distribution
in~\eqref{eq:posterior}, where \(\pi_{\bm \lambda}(\bm \beta)\) is the probability density function
of the improper Gaussian prior for \(\bm \beta\),
\begin{equation}
  \pi_{\bm \lambda}(\bm \beta) \propto \left\lvert\sum_{i = 1}^M \lambda_i \bm S_i\right\rvert^{1/2}_+
  \exp\left(-\frac{1}{2}\bm \beta^\top \left(\sum_{i = 1}^M \lambda_i \bm S_i\right) \bm \beta\right),
\end{equation}
where \(\lvert \bm S\rvert_+\) is the product of all non-zero eigenvalues of \(\bm S\). Instead of
estimating the spline parameters \(\bm \beta\) and the penalty parameters \(\bm \lambda\)
simultaneously, they integrate the spline parameters out of the posterior distribution using a
Laplace approximation to the marginal log-likelihood
\begin{equation}
  \label{eq:marginal_likelihood}
  \mathcal V(\bm \theta, \bm \lambda) = \log \int \exp(\ell(\bm \theta, \bm \beta)) \pi_{\bm \lambda}(\bm
  \beta) \text{d} \bm \beta.
\end{equation}
The SAM model is essentially implemented as a wrapper around the \texttt{TMB} package
\citep{KristensenEtAl2016TMBAutomaticDifferentiation} in \texttt{R}, which contain built-in
functionality for approximating marginal log-likelihood functions using the Laplace approximation.
This makes it straightforward to approximate~\eqref{eq:marginal_likelihood} using Laplace
approximations. Note that the SAM implementation already relies on such Laplace approximations for
integrating out \(N_{a, y}\), \(F_{a, y}\) and several other unknowns from its likelihood.
See the openly available code of \citet{BreivikEtAl2021Predictingabundanceindices} for an example
of how to estimate smoothing penalty parameters in a simplified stock assessment model, using a
Laplace approximation to the marginal log-likelihood in~\eqref{eq:marginal_likelihood}.

Unfortunately, the Laplace approximation does not work well for random effects that are far from
Gaussian distributed \citet[e.g.][]{KristensenEtAl2016TMBAutomaticDifferentiation}. It appears
that this might be the case for some of our random effects, because, when fitting our modified SAM
model to fish stock data from Section~\ref{sec:case-study}, it often fails to converge, when using
the Laplace approximation on the spline parameters in \(\sigma_a^2\) and \(\omega^2_{a, j}\). For
this reason, we choose to not integrate out the variance spline parameters with the Laplace
approximation. Instead, we estimate the penalty and spline parameters for the variance components
simultaneously, using the default optimisation method within SAM. As mentioned in
Section~\ref{sec:model}, simultaneous estimation of penalty and spline parameters is not possible
when optimising over the log-likelihood function. However, the normalisation constant of the added
prior distribution in~\eqref{eq:posterior} rewards larger penalty parameters, and thus makes it
possible to estimate both \(\bm \beta\) and \(\bm \lambda\) simultaneously without the estimator for
\(\bm \lambda\) becoming equal to zero. Note that we still rely on the Laplace approximation for
integrating out all spline parameters related to \(Q_{a, j}\).

In practice, after testing our model on different stock assessment data sets, we find that further
changes should be made to the SAM model to achieve more robust parameter estimation. The
differences between the parameters for the smallest age groups are typically expected to be much
larger than the differences for larger age groups. As described in Section~\ref{sec:model}, we
therefore implement our splines as functions of \(\log(a + 1)\) instead of \(a\). However, in our
experience, this transformation is often not enough to fully account for the differences between
large and small age groups. The need for varying degrees of wiggliness for small and large age
groups can be problematic when estimating a single smoothing penalty parameter for all the age
groups. Thus, we often find that a penalty parameter that is independent of age becomes too strict
for small age groups and too lenient for large age groups. To account for this, we modify the
smoothing penalty by down-weighting it for the youngest age groups. This is achieved by modifying
the penalty matrices \(\bm S_i\) into the matrices
\begin{equation}
\tilde{\bm S}_i = \bm D \bm S_i \bm D.
\end{equation}
Here, the matrix \(\bm D\) is a diagonal matrix with elements that are less than \(1\) for the basis
splines that are centred around the smallest age groups, and \(1\) for all other basis splines. We
set the diagonal elements for the basis splines corresponding to the three youngest age groups equal
to \(\exp(a - 4)\), where \(a \in \{1, 2, 3\}\).

\begin{figure*}
  \centering
  \includegraphics[width=.9\linewidth]{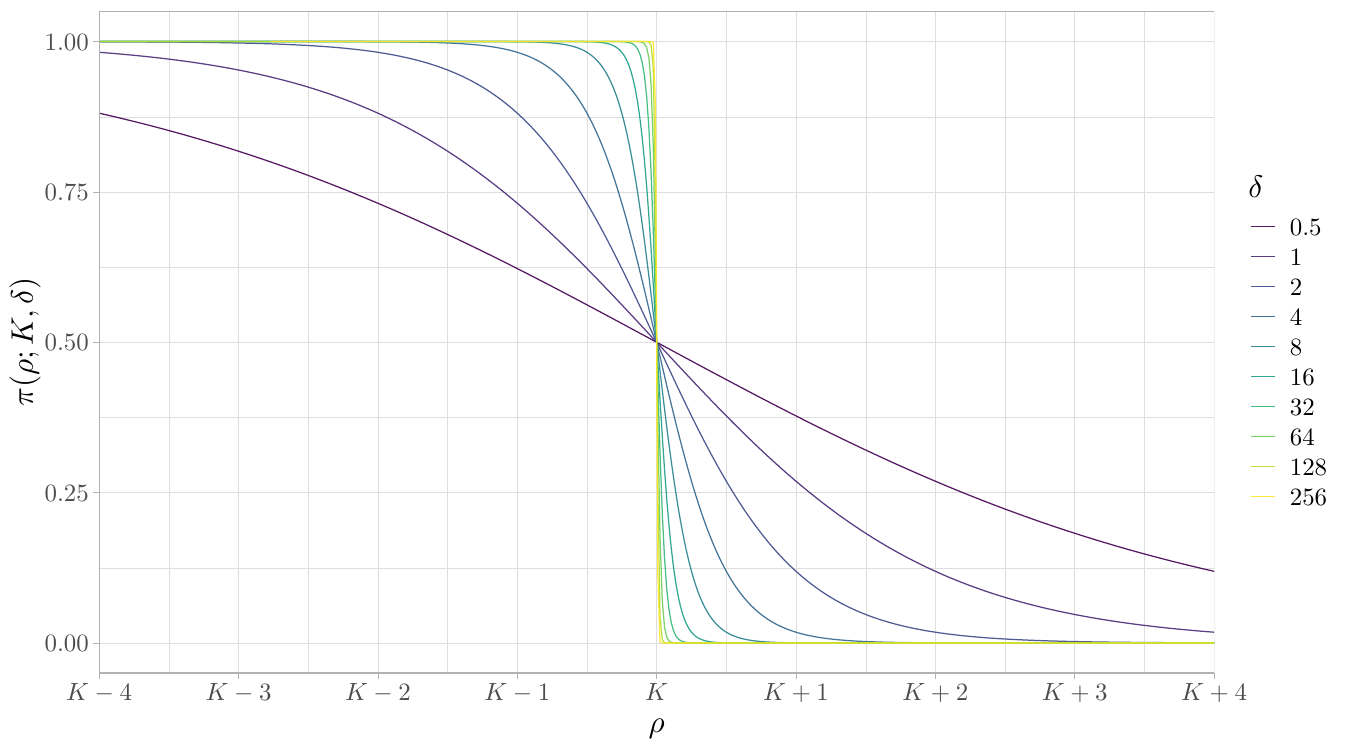}
  \caption{Probability density function of the penalty prior, for a set of different values of
    \(\delta\).}%
  \label{fig:prior}
\end{figure*}

For some stock assessment data sets, we experience weak age signals in \(\sigma^2_{a, I}\).
Sometimes, the age signals are so weak that the estimated splines prefer to be straight lines. This
causes the penalty parameters to diverge towards infinity, which causes numerical problems. To
obtain more numerically robust parameter estimates, we therefore extend the model
in~\eqref{eq:marginal_likelihood} by adding an improper prior distribution to
\(\bm \rho = \log \bm \lambda\) that penalises too large values of \(\bm \rho\). Our chosen prior is
essentially a uniform prior between negative infinity and some large constant \(K\), with a small
probability of allowing larger values than \(K\), in case this becomes necessary. The probability
density function of the improper prior is
\begin{equation}
  \label{eq:penalty_prior}
  \pi(\bm \rho; K, \delta) = \prod_{i = 1}^M \pi(\rho_i; K, \delta) =
  \prod_{i = 1}^M \frac{\exp(\delta (K - \rho_i))}{\left(1 + \exp(\delta (K - \rho_i))\right)},
\end{equation}
where we choose \(K = 7\) and \(\delta = 100\). The parameter \(\delta\) describes how quickly the
prior density should change from \(1\) to \(0\) as \(\rho\) increases and grows larger than
\(K\). Figure~\ref{fig:prior} displays the form of \(\pi(\rho; K, \delta)\) for a set of different
\(\delta\)-values, to visualise its effect on the prior. We choose this penalty prior for multiple
reasons. This improper prior is approximately equal to one for all penalty parameters below
\(\exp(K)\), meaning that it does not affect our parameter estimation unless the penalty parameters
become really large. Furthermore, the prior is smooth, continuous and gives positive probabilities
to all values of \(\bm \rho \in \mathbb R^M\), which is necessary for the automatic differentiation
within \texttt{TMB} to work as it should.

To summarise, after having included all of our changes, the final loss function that we use for
estimating all model parameters, is
\begin{equation}
  \ell(\bm \theta, \bm \beta) + \log \pi_{\bm \lambda}(\bm \beta) + \log \pi(\bm \lambda),
\end{equation}
where \(\ell(\bm \theta, \bm \beta)\) is the log-likelihood of the original SAM-model, modified to
let \(Q_{a, j}\), \(\omega_{a, j}^2\) and \(\sigma_a^2\) be modelled using
spline-functions. Furthermore, the \(\bm \beta\)-prior is an improper Gaussian prior with
probability density function
\begin{equation}
  \pi_{\bm \lambda}(\bm \beta) \propto \left\lvert\sum_{i = 1}^M \lambda_i \bm D \bm S_i \bm D\right\rvert^{1/2}_+
  \exp\left(-\frac{1}{2}\bm \beta^\top \left(\sum_{i = 1}^M \lambda_i \bm D \bm S_i \bm D\right) \bm \beta\right),
\end{equation}
and the improper \(\bm \lambda\)-prior has the probability density function given
in~\eqref{eq:penalty_prior}, with \(K = 7\) and \(\delta = 100\). Then, using \texttt{TMB}, the
spline parameters for \(Q_{a, j}\) are integrated out with the Laplace approximation, along with
several other model parameters that SAM integrates out by default.

\printbibliography%

\end{document}